\documentclass [12pt, letterpaper]{article}
\usepackage{fullpage}
\usepackage{amsmath}
\usepackage{amssymb}
\usepackage{graphicx}
\usepackage{mathrsfs}
\usepackage{wrapfig}
\usepackage{boxedminipage}
\usepackage{epsfig}
\usepackage{setspace}
\usepackage{subfigure}
\usepackage{float}
\usepackage{hyperref}
\newcommand{\reef}[1]{(\ref{#1})}
\begin{document}

\begin{flushright}
\phantom{{\tt arXiv:0709.????}}
\end{flushright}

\bigskip
\bigskip
\bigskip

\begin{center} {\Large \bf Holographic Aspects of  Fermi
    Liquids}
  
  \bigskip

{\Large\bf  in  a }

\bigskip

{\Large\bf Background  Magnetic Field}

\end{center}

\bigskip \bigskip \bigskip \bigskip

\centerline{\bf Tameem Albash, Clifford V. Johnson}

\bigskip
\bigskip
\bigskip

  \centerline{\it Department of Physics and Astronomy }
\centerline{\it University of
Southern California}
\centerline{\it Los Angeles, CA 90089-0484, U.S.A.}

\bigskip

\centerline{\small \tt talbash,  johnson1,  [at] usc.edu}

\bigskip
\bigskip


\begin{abstract} 
\noindent 
We study the effects of an external magnetic field on the properties
of the quasiparticle spectrum of the class of 2+1 dimensional strongly
coupled theories holographically dual to charged AdS$_4$ black holes
at zero temperature. We uncover several interesting features.  At
certain values of the magnetic field, there are multiple quasiparticle
peaks representing a novel level structure of the associated Fermi
surfaces. Furthermore, increasing magnetic field deforms the
dispersion characteristics of the quasiparticle peaks from non--Landau
toward Landau behaviour. At a certain value of the magnetic field,
just at the onset of Landau--like behaviour of the Fermi liquid, the
quasiparticles and Fermi surface disappear.
\end{abstract}
\newpage \baselineskip=18pt \setcounter{footnote}{0}


\section{Introduction}

With increasing interest in the various long--anticipated potential
applications to condensed matter and nuclear physics of gauge/gravity
duality techniques such as AdS/CFT and its
generalisations\cite{Maldacena:1997re,Witten:1998qj,Gubser:1998bc,Witten:1998zw},
it has been of considerable interest to understand how to cleanly
capture the physics of the Fermi surface of strongly coupled systems
using holographic methods. This is because of the potential utility
the methods might have to complement and extend the already impressive
and ubiquitous power of the Fermi liquid paradigm, and also because
there is a wealth of experimental phenomena that suggest that there
are Fermi liquids that have behaviour that lies beyond the reach of
standard techniques. Holographic methods may allow access to such
behaviour.

Recently, the work of ref.\cite{Liu:2009dm} presented convincing
evidence for the Fermi surface of the strongly interacting 2+1
dimensional system at zero temperature and finite density that is
holographically dual\cite{Chamblin:1999tk,Chamblin:1999hg} to an
extremal electrically charged Reissner--Nordstr\"{o}m black hole in
four dimensional anti--de Sitter spacetime (AdS$_4$). (That work
continued and refined the important initial studies presented in
ref.\cite{Lee:2008xf}, and there has been further discussion and
extension of this particular line of development in
refs.\cite{Cubrovic:2009ye,Faulkner:2009wj}. See also
ref.\cite{Rozali:2008jx} for a discussion of possible holographic
descriptions of Fermi surfaces in a different context.)

A key observation of ref.\cite{Liu:2009dm} was the fact that the Fermi
liquid had distinctly non--Landau behaviour, showing unusual
dispersion characteristics of the quasiparticle peak. This is perhaps
to be expected, not just because the system is strongly coupled
(remember that Landau Fermi--liquids arise in strongly coupled
contexts too), but because one might anticipate that the effectively
weakly coupled quasiparticle theory that arises (after strong coupling
dressing) in the vicinity of the Fermi surface ought not to have a
(simple) gravitational dual. When all is said and done, it is, after
all, just a free field theory\footnote{See, however, the work of
  ref.\cite{Cubrovic:2009ye}, but study it alongside the discussion in
  section VI.E. of ref.\cite{Faulkner:2009wj}.}. One might wonder,
however, what range of dispersion characteristics might be accessible
using holographic duals, and in particular how close one can get to
Landau--like behaviour for a given system.  The recent work of
ref.\cite{Faulkner:2009wj} answers some of this by doing a careful
analysis of the AdS$_2\times \mathbb{R}^2$ throat region region (first
discussed in this holographic context in ref.\cite{Chamblin:1999tk})
that appears near the horizon of the black hole at zero
temperature. Much of the physics of the critical exponents can be
traced to the IR physics located down this throat and is controlled by
the masses of fields there.

Our work shows a different way of deforming the critical behaviour. As
we will show, adding a magnetic field moves the dispersion of the
quasiparticle spectrum back {\it toward} Landau--like behaviour, in a
manner that may have precise experimental analogues, especially given
how natural magnetic fields are as laboratory probes and control
parameters in a condensed matter context.  Moreover, we find that at a
given value of the magnetic field ${\cal H}$ (below what appears to be
a certain limiting value ${\cal H}_{\rm max}$), multiple quasiparticle
peaks can appear in the spectrum, representing a {\it finite} series
of levels (a kind of ``band'' structure).  Our boundary condition is such that we are at the zeroth Landau level, so the levels we see are not the infinite family of equally spaced Landau--Rabi levels known from weak coupling intuition.

To introduce a background magnetic field ${\cal H}$ we simply add
magnetic charge to the AdS$_4$--RN black hole, making it a dyon. We
then study the system by probing it (as done in the ${\cal H}=0$ case)
with a minimally coupled Dirac fermion of electric charge $q$. We
compute the retarded Green's function associated with this probe, from
which, after Fourier transforming to momentum--frequency space
$(k,w)$, we extract our physics. 

Generically, it can be seen that the magnetic field shifts the
effective mass of the fermion, which translates into a deformation of
effective dimension of the operator we are probing the theory
with. Following the discussion in
refs.\cite{Cubrovic:2009ye,Faulkner:2009wj}, it is not hard to infer
that this will certainly affect the dispersion of the resulting
quasiparticle peaks, but this needs to be explored explicitly, and
this is what we report on in this paper.

\section{Background}
The metric for the dyonic black hole in asymptotically AdS$_4$
spacetime, using Cartesian coordinates is\cite{Romans:1991nq}:
\begin{eqnarray} \label{eqt:dyonic_metric}
ds^2 &=& \frac{L^2 \alpha^2}{z^2} \left( - f \left(z \right) dt^2 + dx^2 + d y^2 \right) +\frac{L^2}{z^2} \frac{d z^2}{f \left(z \right)} \ ,  \\
F &=& 2 H \alpha^2  dx \wedge d y + 2 Q \alpha dz \wedge dt\ , \nonumber \\
f \left(z \right) &=& 1 + \left( H^2 + Q^2 \right) z^4 - \left(1 + H^2 + Q^2 \right) z^3 = \left(1-z \right)\left( z^2 +z+ 1 - \left(H^2 +Q^2\right) z^3 \right) \ . \nonumber
\end{eqnarray}
Here $L$ is the length scale set by the negative cosmological constant
$\Lambda=-3/L^2$ in Einstein--Maxwell theory given by:
\begin{eqnarray} \label{eqt:EM_action}
S_{\mathrm{bulk}} = \frac{1}{2 \kappa_4^2}  \int d^4 x \sqrt{-G} \biggl\{ R + 
\frac{6}{L^2}  -\frac{L^2}{4} F^2  \biggr\}\ , 
\end{eqnarray}
where $\kappa_4^2=8\pi G_{\rm N}$ is the gravitational coupling and
our signature is $(-+++)$.  The mass per
unit volume and temperature of the black hole are:
\begin{equation}
\label{eqn:massandtemperature}
{\varepsilon}=\frac{\alpha^3L^2}{\kappa^2_4}[1+Q^2+H^2]\ ,\quad
T=\frac{\alpha}{4\pi}[3-(Q^2+H^2)]\ .
\end{equation}
These are also the energy density and temperature of the 2+1 dual
theory, which can be roughly thought of as living on the boundary at
$z=0$.  The horizon of the hole is at $z=1$.  We choose our gauge such
that the gauge field can be written in the following form:
\begin{equation}
A_t = 2 Q  \alpha (z-1) \ , \quad A_x = - 2 H \alpha^2 y \ .
\end{equation}
The electric component sets, by virtue of its value on the boundary, a
chemical potential $\mu=-2Q\alpha$ for the $U(1)$ charge, while the
magnetic component determines a background magnetic field of magnitude
${\cal H}=-2H\alpha^2$.The parameter $\alpha$ has dimensions of
inverse length.  For simplicity, in the sequel we make the following
redefinitions so that we are working entirely in terms of
dimensionless fields and coordinates:
\begin{equation}
t \to t / \alpha \ , \quad x \to x / \alpha \ , \quad y \to y / \alpha \ , \quad A_t \to \alpha A_t \ .
\end{equation}

\subsection{The Probe Fermion}

We consider the Dirac action in this background:
\begin{equation}
S_{D} = \int d^4 x \sqrt{-G} \  i \left( \bar{\Psi} \Gamma^M \mathcal{D}_M \Psi - m \bar{\Psi} \Psi \right) \ , 
\end{equation}
where $\mathcal{D}_M$ is the covariant derivative given by:
\begin{equation}
\mathcal{D}_M = \partial_M + \frac{1}{4} \omega_{a b M} \Gamma^{a b} - i q A_M \ ,
\end{equation}
and $M$ are world indices and $a,b$ are tangent--space indices.  $\omega_{a b M}$ is the spin connection given by:
\begin{equation}
\omega_{a b M} = e_a^N \partial_M e_{b N} - e_{ a N} e_{b}^O \Gamma^{N}_{O M} \ ,
\end{equation}
and
\begin{equation}
\Gamma^{a b} = \frac{1}{2} \left[\Gamma^a, \Gamma^b \right] \ , \Gamma^{M} = e^{M}_a \Gamma^a \ .
\end{equation}
The Dirac equation is given by:
\begin{equation}
\Gamma^M \mathcal{D}_M \Psi - m \Psi = 0 \ .
\end{equation}
We choose $\Psi$ such that:
\begin{equation}
\Psi =  z^{3/2} f^{-1/4} e^{- i \omega t + k_x x} \left(\begin{array}{c}
\phi_+ (y,z) \\
\phi_- (y,z) 
\end{array}
\right) \ ,
\end{equation}
and
\begin{equation}
\Gamma^3 = \left( \begin{array}{cc}
1 & 0 \\
0 & -1
\end{array} \right) \ , \quad \Gamma^\mu = \left( \begin{array}{cc}
0 & \gamma^\mu \\
\gamma^{\mu} &  0
\end{array} \right) \ .
\end{equation}
This reduces the equation of motion to:
\begin{eqnarray}
\sqrt{ \frac{g_{xx}}{g_{zz}}} \left( \partial_z  \mp m \sqrt{g_{zz}}  \right) \phi_\pm &=& \pm i \left( \gamma^0 u  + i \gamma^2  \partial_y   - \gamma^1 \left( 2 H q y + k_x  \right)  \right) \phi_\mp
 \end{eqnarray}
where:
\begin{equation}
u = \sqrt{ \frac{g_{xx}}{-g_{tt}}} \left( \omega + 2 q Q (z-1) \right) \ , \ \sqrt{ \frac{g_{xx}}{g_{zz}}} = \sqrt{f} \ ,  \sqrt{ \frac{g_{xx}}{-g_{tt}}} = \frac{1}{\sqrt{f}} \ .
\end{equation}
Consider the limit of $z \to 0$ of this equation.  The solution is given by:
\begin{eqnarray}
\phi_+ &=& A z^m+ z^{1-m} \frac{  i \left( \gamma^0 u - \gamma^1 \left(2 H q y + k_x \right) + i \gamma^2 \partial_y \right)}{1 - 2m } B \ , \\
\phi_- &=& B z^{-m} + z^{m+1} \frac{-i \left(\gamma^0 u + \gamma^1 \left(2 H q y + k_x \right) - i \gamma^2 \partial_y \right)}{1 + 2m } A \ ,
\end{eqnarray}
where $A$ and $B$ are functions independent of $z$.  The various
relationships between the various powers of $z$ is relevant for the
AdS/CFT dictionary \cite{Iqbal:2009fd}.  In particular, if we restrict
ourselves to $m \geq 0$, then the source of the dual operator is
proportional to $B$, and the vacuum expectation value (vev) of the
operator is proportional to $A$.  In this case, the dimension of the
operator is given by:
\begin{equation}
\Delta = \frac{d}{2} + m \ . 
\end{equation}
We now proceed away from $z=0$ and further decompose the two--component fields $\phi_\pm$ as:
\begin{equation}
\phi_\pm = \left( \begin{array} {c}
\chi_\pm \\
\xi_\pm
\end{array} \right) \ ,
\end{equation}
and make the following choices for the $\gamma$'s:
\begin{equation}
\gamma^0 = i \sigma_2 \ , \quad \gamma^1 = \sigma_1 \ , \quad \gamma^2 = \sigma_3 \ ,
\end{equation}
and these fields satisfy the following coupled equations of motion
\begin{eqnarray}
\sqrt{ \frac{g_{xx}}{g_{zz}}} \left( \partial_z  \mp m \sqrt{g_{zz}} \right) \chi_\pm & =& \mp  \left[ - i  u  \xi_\mp + \partial_{y} \chi_\mp + i \left( 2 H q y + k_x \right) \xi_\mp \right] \ , \\
\sqrt{ \frac{g_{xx}}{g_{zz}}} \left( \partial_z  \mp m \sqrt{g_{zz}} \right) \xi_\pm & =& \mp  \left[ i  u \chi_\mp - \partial_{y} \xi_\mp + i \left( 2 H q y + k_x \right) \chi_\mp \right] \ .
\end{eqnarray}

\section{Specializations}{
  \subsection{Zero Temperature}
Henceforth we will work at zero temperature. This means, from
equation~\reef{eqn:massandtemperature} that there is a relation between $Q$
and $H$ that we must bear in mind:
\begin{equation}
Q=\left(3-H^2\right)^{\frac12}\ .
\end{equation}
In particular, note that (we temporarily restore $\alpha$ for this
discussion) the chemical potential $\mu=-2\alpha Q$ naively gets
shifted from its $(T=0,H=0)$ value due to this relation.  Also of note
is the fact that as $H$ increases, $Q$ must decrease in order to keep
the temperature vanishing. There is a maximum value of $H$, $H_{\rm
  max}=\sqrt{3}$, at which $Q$ vanishes.  This is all to be physically
interpreted for our system as follows. First, we need to decide what
physical quantity we are holding fixed while adding magnetic field to
the system. A good such quantity is the chemical potential
$\mu=-2\alpha Q$. Even though $Q$ decreases for non--zero $H$, we
simply increase $\alpha$ to hold it at the same value that it was at
zero $H$.(According to equation~\reef{eqn:massandtemperature}, this
increases the energy density $\varepsilon$ of the field theory, but
this is quite natural as the magnetic field lifts the available energy
levels of all charged particles.) Writing $3-H^2={\tilde Q}^2
\delta^2$ we can write $\alpha={\tilde \alpha}\delta^{-1}$ so that
$\mu$ is fixed as $\delta\to0$. Meanwhile we see that the physical
magnetic field ${\cal H}=-2\alpha^2 H$ goes to infinity, showing that we do
not have a limiting {\em physical} applied  magnetic field.

\subsection{A Zero Temperature Ansatz}

In order to satisfy the appropriate boundary conditions at the event
horizon, we redefine the fields as follows (for $\omega\neq0$; see
subsection~\ref{sec:omegazero} for a comment on the $\omega=0$ case):
\begin{eqnarray}  \label{eqt:field_redef_1}
\chi_\pm &=& a_\pm (y, z) \exp \left( \frac{i \omega}{6 \left(1-z \right)} \right) \left(1-z \right)^{i \left( 6 q Q - 4 \omega \right)/18 } \ , \nonumber \\
\xi_\pm &=& b_\pm (y, z) \exp \left( \frac{i \omega}{6 \left(1-z \right)} \right) \left(1-z \right)^{i \left( 6 q Q - 4 \omega \right)/18 } \ .
\end{eqnarray}
With this field redefinition, the equations of motion expanded at the event horizon give the following conditions:
\begin{equation} \label{eqt:eh_cond1}
a_+ (y, 1) = b_- (y,1) \ , \quad a_- (y,1) = - b_+(y, 1) \ , 
\end{equation}
This result suggests that the ``correct'' variables to study the problem are actually:
\begin{equation}
A_+ (y,z) = b_-(y,z) - a_+(y,z) \ , \quad A_- (y,z)= - i \left(a_-(y,z) + b_+(y,z) \right) \ , 
\end{equation}
\begin{equation}
B_+(y,z) = a_+(y,z) + b_- (y,z) \ , \quad B_- (y,z)= i \left(b_+(y,z) - a_- (y,z) \right) \ .
\end{equation}
The equations of motion are now given by:
\begin{eqnarray}
\sqrt{\frac{g_{xx}}{g_{zz}}} \left(\partial_z  + \frac{ i \omega}{6 (1-z)^2}   + i \frac{- 6 q Q + 4 \omega}{18 (1 - z)}  \right) A_+&=&\\
&&\hskip-3cm-\, \sqrt{g_{xx}} m B_+ - i u A_+ + i \left( \partial_y B_- + (2H q y + k_x) B_+ \right) \ , \nonumber\\
\sqrt{\frac{g_{xx}}{g_{zz}}} \left( \partial_z  + \frac{ i \omega}{6 (1-z)^2}   + i \frac{- 6 q Q + 4 \omega}{18 (1 - z)}  \right) A_-&=&\\  
&&\hskip-3cm-\, \sqrt{g_{xx}} m B_- - i u A_-  - i \left( \partial_y B_+ + (2H q y + k_x) B_- \right) \ , \nonumber\\
\sqrt{\frac{g_{xx}}{g_{zz}}} \left( \partial_z  + \frac{ i \omega}{6 (1-z)^2}   + i \frac{- 6 q Q + 4 \omega}{18 (1 - z)}  \right) B_+ &=& \\  
&&\hskip-3cm-\, \sqrt{g_{xx}} m A_+ + i u B_+ -i  \left( \partial_y A_- + (2H q y + k_x) A_+ \right) \ , \nonumber \\
\sqrt{\frac{g_{xx}}{g_{zz}}} \left( \partial_z  + \frac{ i \omega}{6 (1-z)^2}   + i \frac{- 6 q Q + 4 \omega}{18 (1 - z)}  \right) B_- &=& \\  
&&\hskip-3cm-\, \sqrt{g_{xx}} m A_- + i u B_- + i \left( \partial_y A_+ + (2H q x + k_x) A_- \right) \ . \nonumber
\end{eqnarray}
At the event horizon, we impose the following condition:
\begin{equation} \label{eqt:eh_condition}
\partial_y B_\pm (y,1) + 2 H q y B_\mp (x,1) = 0 \ , 
\end{equation}
which has solution given by:
\begin{equation} \begin{array}{c}
B_+ (y,1) = B_- (y,1) = B_0 \exp \left( -  H q y^2 \right) \ , \\
B_+ (y,1) =  - B_- (y,1) = B_0 \exp \left( H q y^2 \right) \ ,
\end{array}
\end{equation}
%
This solution can be understood as an infinite sum of the separable solutions of the problem in terms of the coordinate $ \eta = \sqrt{ 2 H q} \left( y + k_x / 2 H q \right)$.  For this reason, we refer to this solution as the ``infinite--sum'' solution. We elaborate more on this relationship in our companion paper~\cite{Albash:2010yr}.
%
This boundary condition restricts us to the zeroth Landau level.  This then leaves us with the following conditions at the event horizon:
\begin{equation}
A_\pm (r,1) = 0 \ , \quad \partial_z A_\pm (r,1) = \mp \frac{\sqrt{6}}{2 \omega} \left(k_x  \pm i m \right) B_\pm \ ,
\end{equation}
\begin{eqnarray}
\partial_z B_\pm (x,1) &=& - \frac{i}{108} \left(\left( 48 q Q - 23 \omega \right) B_\pm \mp 18 \sqrt{6}  \left(k_x \mp i m \right) \partial_z A_\pm(x,1) \right. \nonumber \\
&& \left. \mp 18 \sqrt{6} \partial_z \left( \partial_y A_\mp + 2 H q y A_\pm \right) \right) \ , \nonumber \\
&=& - \frac{i}{108} \left(48 q Q - 23 \omega  + \frac{54}{\omega}  \left(k_x^2 + m^2  + 4 H q y \left( k_x \mp i m \right) \right) \right) B_0 \ .
\end{eqnarray}
Which solution is chosen depends on the direction of the magnetic
field.  For the rest of our analysis, we stick to the solution with
$B_+ = B_-$ which means we have negative $H$.

In all that follows, we restrict ourselves to $m=0$ for our probe
fermion.  By the AdS/CFT dictionary, this means we are turning on
operators with dimension $\Delta = 3/2$.  This case was studied in
great detail for the case of $H=0$ in ref.~\cite{Liu:2009dm}.
\subsection{Extracting the retarded Green's function}
%
Following the prescription for extracting the Green's function from
ref.~\cite{Iqbal:2009fd}, the Green's function is given by (assuming $m
\geq 0$):
\begin{equation}
G_R = \lim_{\epsilon \to 0} \epsilon^{-2m} \left( \begin{array}{cc}
i \frac{\chi_+(k)}{\xi_-(k)} & 0 \\
0 & - i \frac{\xi_+(k)}{\chi_-(k)}
\end{array} \right) \ ,
\end{equation}
which in term of our fields $A_\pm$ and $B_\pm$ gives:
\begin{equation}
G_+=i \frac{\chi_+(k)}{\xi_-(k)} =i \frac{B_+(k) - A_+(k)}{B_+(k) + A_+(k)} \, \quad G_-=  -i \frac{\xi_+(k)}{\chi_-(k)}  = i \frac{B_-(k) - A_-(k)}{B_-(k) + A_-(k)}  \ .
\end{equation}
Note that these fields are given in momentum--space and not
position--space.  A subtlety here is that in ref.~\cite{Iqbal:2009fd},
the fields always had a separable form with respect to the AdS radial
coordinate and the space--time coordinates of the field theory,
whereas here this is not the case.  We assume that the prescription
continues to hold and we simply Fourier transform our fields at the
AdS boundary and use the result in the prescription.
%

%
\subsection{The case of $\omega = 0$.}
%
\label{sec:omegazero}
For the special case of $\omega = 0$, the condition $A_\pm (y,1)= 0 $
is no longer required.  In fact, we need to change the behavior of the
fields at the event horizon.  Consider first  the $H=0$ case.  We go
back to equations~\reef{eqt:field_redef_1} and make the following
redefinition:
\begin{eqnarray}
\chi_\pm(z) &=& a_\pm(z) \left(1-z \right)^{\frac{\sqrt{2}}{6} \sqrt{3 k^2 - 2 q^2 Q^2}} \ , \\
\xi_\pm(z) &=& b_\pm(z) \left(1-z \right)^{\frac{\sqrt{2}}{6} \sqrt{3 k^2 - 2 q^2 Q^2}} \ .
\end{eqnarray}
The exponent of the $(1-z)$ term is chosen such that there is a
non--zero solution for the fields.  Now we define $A_\pm$ and $B_\pm$
as before, and find the following condition at the event horizon:
\begin{eqnarray}
A_+ (1) &=& - \frac{1}{\sqrt{6}k} \left( 2 q Q + i \sqrt{2} \sqrt{3 k^2 - 2 q^2 Q^2} \right) B_+(1) \ , \\
A_- (1) &=&  \frac{1}{\sqrt{6}k} \left( 2 q Q + i \sqrt{2} \sqrt{3 k^2 - 2 q^2 Q^2} \right) B_-(1) \ .
\end{eqnarray}
In particular, if we take $B_+(1) = B_-(1)$, we have $A_-(1) = -
A_+(1)$, and we get for our Green's functions at the event horizon:
\begin{equation}
G_+ = - \frac{\sqrt{3 k^2 - 2 q^2 Q^2}}{\sqrt{3} k - \sqrt{2} q Q} \ , \quad G_- = \frac{\sqrt{3 k^2 - 2 q^2 Q^2}}{\sqrt{3} k + \sqrt{2} q Q} \ .
\end{equation}
Another independent solution is to take:
\begin{eqnarray}
\chi_\pm(z) &=& a_\pm(z) \left(1-z \right)^{-\frac{\sqrt{2}}{6} \sqrt{3 k^2 - 2 q^2 Q^2}} \ , \\
\xi_\pm(z) &=& b_\pm(z) \left(1-z \right)^{-\frac{\sqrt{2}}{6} \sqrt{3 k^2 - 2 q^2 Q^2}} \ ,
\end{eqnarray}
which gives at the event horizon:
\begin{eqnarray}
A_+ (1) &=& - \frac{1}{\sqrt{6}k} \left( 2 q Q - i \sqrt{2} \sqrt{3 k^2 - 2 q^2 Q^2} \right) B_+(1) \ , \\
A_- (1) &=&  \frac{1}{\sqrt{6}k} \left( 2 q Q - i \sqrt{2} \sqrt{3 k^2 - 2 q^2 Q^2} \right) B_-(1) \ .
\end{eqnarray}
Again if we take $B_+(1) = B_-(1)$, we have $A_-(1) = -
A_+(1)$  but with the Green's function gives as:
\begin{equation}
G_+ =  \frac{\sqrt{3 k^2 - 2 q^2 Q^2}}{\sqrt{3} k - \sqrt{2} q Q} \ , \quad G_- = - \frac{\sqrt{3 k^2 - 2 q^2 Q^2}}{\sqrt{3} k + \sqrt{2} q Q} \ .
\end{equation}
If we now turn on $H$, the situation becomes much more
complicated. The equations of motion for $A_\pm$ and $B_\pm$ at the
horizon require a more complicated exponent for $(1-z)$ in their
behaviour. For our purposes, we will have no need to focus on the case
of $\omega=0$ and determine these exponents since the interesting
physics (specifically, the quasiparticle peaks) will appear away from
$\omega=0$. We will therefore not pursue this further in the present
work.

\section{Solutions}

\subsection{A Separable Solution}
Recalling that we will work with $m=0$ henceforth, we begin by
considering the trivial case of a separable solution which occurs for
$k_x=0$, {\it i.e.}
\begin{equation} \label{eqt:ansatz_separable}
A_\pm (y,z)= A(z) \exp \left ( - H q y^2  \right) \ , \quad B_\pm (y,z)= B(z) \exp \left (-  H q y^2  \right) \ .
\end{equation}
In this case, the four equations of motion reduce to two:
\begin{eqnarray}
\sqrt{\frac{g_{xx}}{g_{zz}}} \left(\partial_z A + \frac{ i \omega}{6 (1-z)^2} A  + i \frac{- 6 q Q + 4 \omega}{18 (1 - z)} A \right)&=& - i u A \ , \\
\sqrt{\frac{g_{xx}}{g_{zz}}} \left( \partial_z B + \frac{ i \omega}{6 (1-z)^2} B  + i \frac{- 6 q Q + 4 \omega}{18 (1 - z)} B \right)&=&  i u B \ .
\end{eqnarray}
Therefore in the separable case, $G_R(1,1) = G_R(2,2)$.  In fact,
looking at the initial conditions at the event horizon for $A(z)$, we
find that $\partial_z A(1) = A(1) = 0$, and therefore we can
consistently take $A(z) = 1$.  Therefore, we automatically get that
$G_R(1,1) = i$.
\\
The other separable solution occurs for $H=0$, and our solution should
match those of ref.~\cite{Liu:2009dm}.  In this case, we can take:
\begin{equation}
A_\pm (y,z) = A_\pm(z) \ , \quad B_\pm(y,z) = B_\pm(z)  \ .
\end{equation}
Choosing $q=-0.5$ we do indeed recover the results of
ref.~\cite{Liu:2009dm}, as we show in figure \ref{fig:H=0}.
\begin{figure}[h] 
   \centering
   \includegraphics[width=2in]{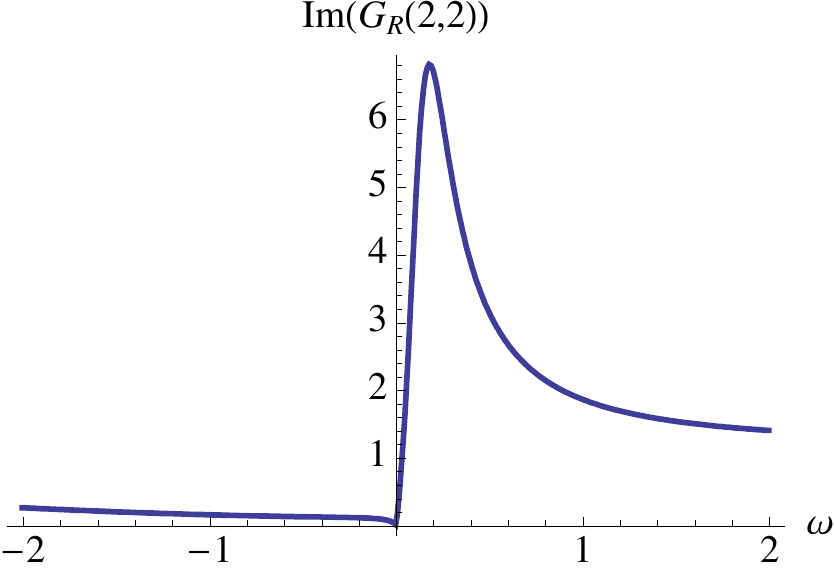} 
   \caption{\small Solution for $H=0$ and $k_x = 1.2$ using Mathematica's NDSolve}
   \label{fig:H=0}
\end{figure}
These two cases of separable solutions provide good test cases for the
accuracy of numerical procedure when we solve the infinite--sum cases
next.

Before proceeding, we emphasize that our separable ansatz in equation \reef{eqt:ansatz_separable} only considers the zeroth order Hermite function. The ansatz can be generalized for higher order Hermite functions, the physics of which we study in a companion paper \cite{Albash:2010yr}.
%
\subsection{Infinite--sum cases: Full Numerical Method}
%
We now proceed to solve the problem with $k \neq 0 $ and $H \neq 0$.
We choose to discretize the equations of motion using a
``fully--implicit'' method:
\begin{eqnarray}
A_+( y_j, z_n) &=& A_+ ( y_j, z_{n-1}) - i \Delta z F_+(z_{n-1}) A_+(y_j,z_{n-1})  \\
&& \hskip-1cm + i \frac{\Delta z}{\sqrt{f(z_{n-1})}} \left( \frac{B_- ( y_{j+1}, z_{n-1}) - B_- (y_{j-1}, z_{n-1})}{2 \Delta y} + \left(2 H q y_j + k_x \right) B_+(y_j,z_{n-1})\right) \ , \nonumber
\end{eqnarray}
where we have defined:
 \begin{equation}
 F_+ (z) =   \frac{\omega}{6 (1-z)^2} + \frac{ - 6 q Q + 4 \omega}{18(1-z)} + \frac{u}{\sqrt{f(z)}} \ . 
 \end{equation}
 The problem is now to solve a large number of linear equations at
 each $z$ step.  The associated matrix is of block tri--diagonal form,
 i.e. :
 \begin{equation}
\left( \begin{array}{cccccccc}
 \ddots & &&&&&& \\
 \cdots & A_1 & B_1 & C_1 & 0&0&\cdots \\
\cdots & 0 &A_2 & B_2 & C_2 &0& \cdots \\
\cdots &  0 &0 &A_3 & B_3 & C_3 & \cdots \\
 & &&&&& \ddots
 \end{array}\right) \ ,
  \end{equation}
  where each of the $A_i, B_i, C_i$'s are matrices ($8\times 8$
  matrices in this case).  This matrix can be inverted using a
  ``forward elimination--backward substitution'' method.
 %
 \subsubsection{Accuracy of Numerics:  Two  Tests}
 %
 The first test of our numerical method is to reproduce the known
 separable results we describe above.  We present the result for
 $H=0$, (with $k_y=0$ and non--zero $k_x$) in figures
 \ref{fig:H=0_c++} and~\ref{fig:H=0_scaling}, demonstrating with the
 latter that we can reproduce the correct quasiparticle peak and its
 associate scaling behaviour.
 \begin{figure}[ht]
\begin{center}
\subfigure[$k_x=1.2$]{\includegraphics[width=2.5in]{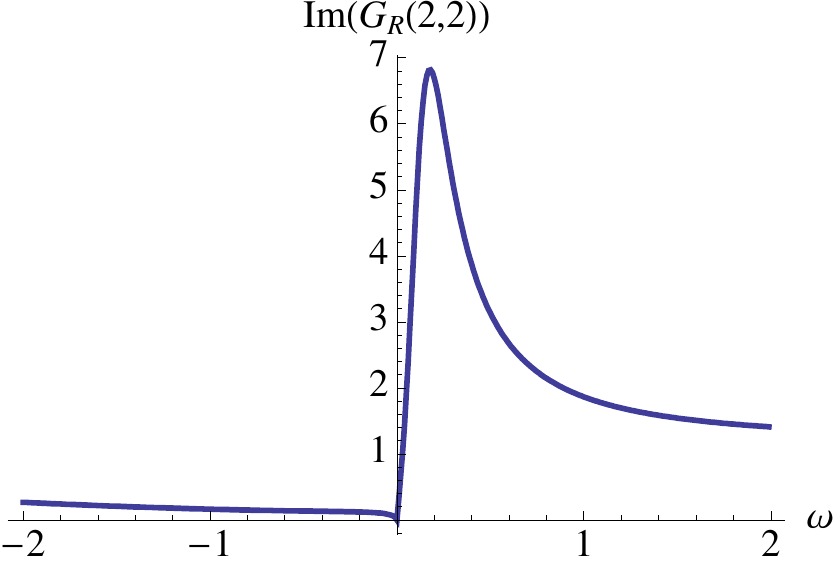}\label{fig:H=0_kx=1}} \hspace{0.5cm}
\subfigure[$k_x=3$]{\includegraphics[width=2.5in]{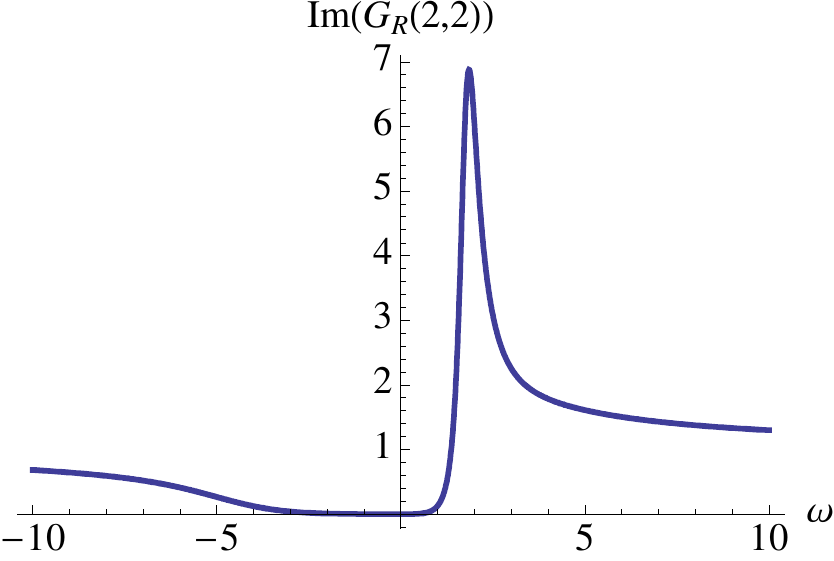}\label{fig:H=0_kx=3}} 
   \caption{\small Solution for $H=0$ using implicit method to solve PDE.}  \label{fig:H=0_c++}
   \end{center}
\end{figure}
 \begin{figure}[ht]
\begin{center}
\subfigure[$\omega$ scaling]{\includegraphics[width=2.5in]{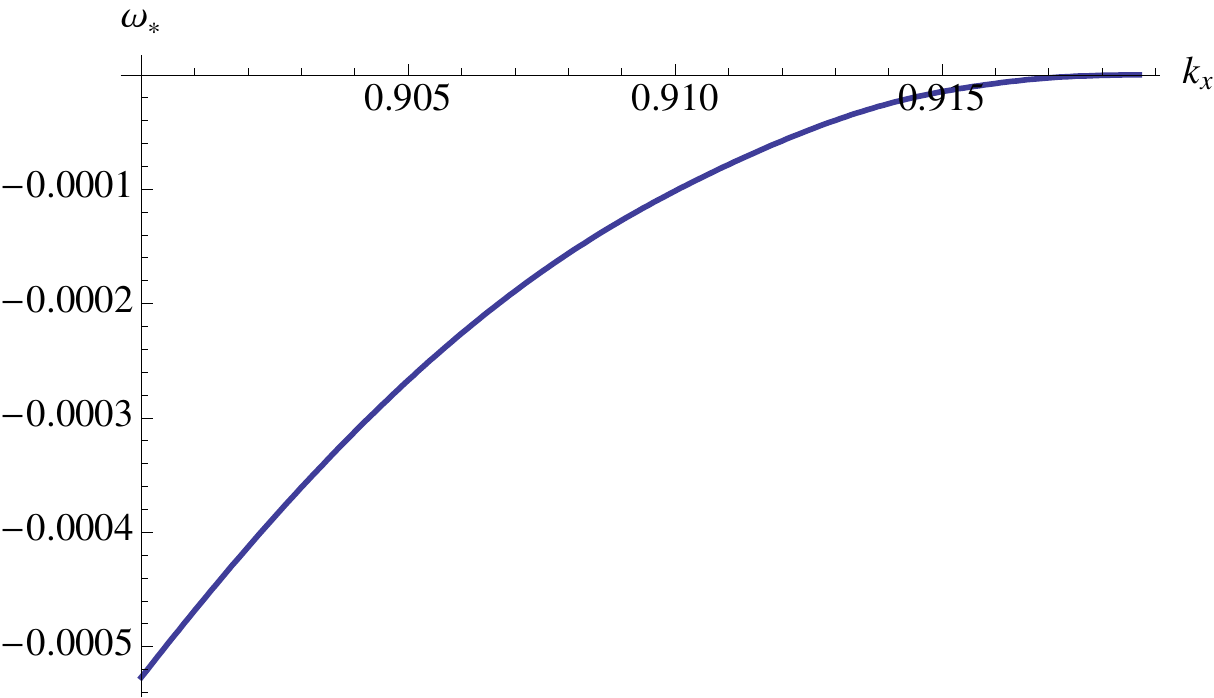}\label{fig:scaling_w_H=0}} \hspace{0.5cm}
\subfigure[Peak scaling]{\includegraphics[width=2.5in]{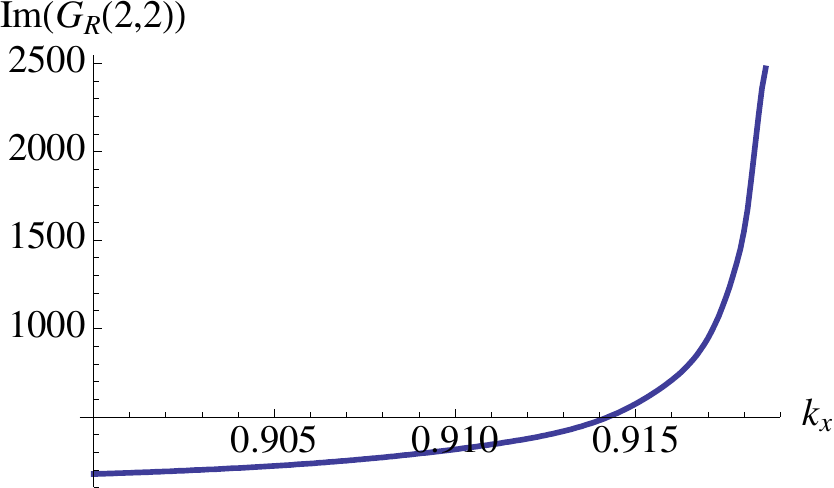}\label{fig:scaling_G_H=0}} 
   \caption{\small Scaling behavior of $\omega^*$ (location) and the height of the peak near the pole at $\omega = 0$ and $k_x \approx 0.918$.}  \label{fig:H=0_scaling}
   \end{center}
\end{figure}
It is interesting to test the case of $k_x= 0$. 
 
We find considerable deviation from the expected answer of $G_R(2,2) =
i$.  We can understand this completely as arising from numerical
error. The field $A_+$ at the boundary should be zero in this case,
but it turns out to be proportional to the derivative of the $B$'s.
It is straight--forward to understand this discrepancy: it comes from
the finite difference approximation to the $y$--derivative, which has
zeros for non--zero $H$ at zero $k_x$, causing a loss of accuracy.
Given that we have rotational symmetry, we can choose, without loss of
generality, to put all our momentum into $k_x$ and work in the more
well--behaved $k_y=0$ sector for the remainder of our investigations.

We are now ready to numerically explore $(k,\omega)$ landscape for
various $H$, searching for quasiparticle poles in the Green's
function. 

\section{Observations}

There are several important general features that we encountered in
our numerical explorations, and we collect them all together here.

\subsection{A Gap, A Ridge, and Some Poles}
\begin{itemize}
\item For a given magnetic field, as $k_x$ is increased, a region
  appears where Im($G_R(2,2)$) is negative for values of $\omega <
  \omega_{\mathrm{gap}}$.  We explain the meaning of this region
  below, but we illustrate in figure \ref{fig:gapbehavior} how
  $\omega_{\mathrm{gap}}$ behaves as we increase $k_x$ for fixed
  magnetic field.  In particular, we note that as $k_x \to \infty$,
  the behavior of $\omega_{\mathrm{gap}}$ is linear and appears to
  remain a fixed distance from the $\omega = k_x$ line.  For higher
  magnetic fields, the separation from the line decreases.
\begin{figure}[ht]
\begin{center}
\subfigure[$H=-0.1$]{\includegraphics[width=2.5in]{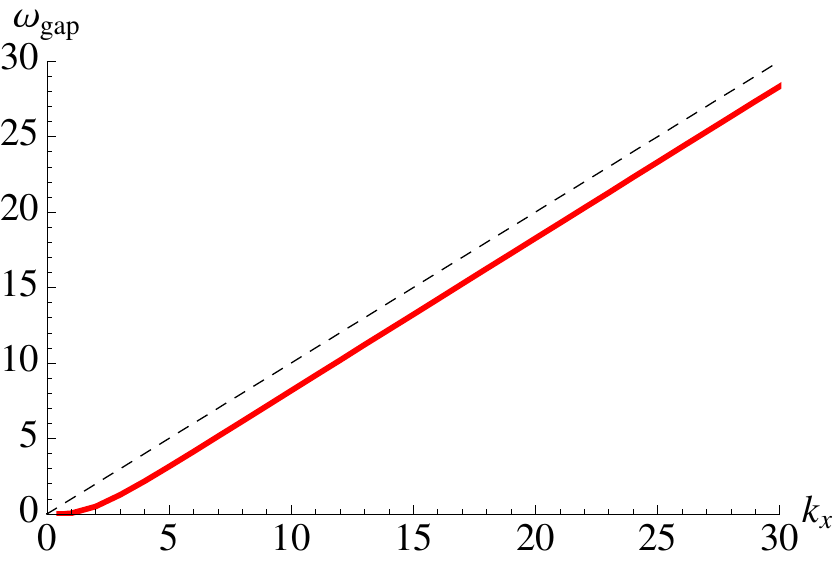}\label{fig:wgap0.1}} \hspace{0.5cm}
\subfigure[$H=-0.35$]{\includegraphics[width=2.5in]{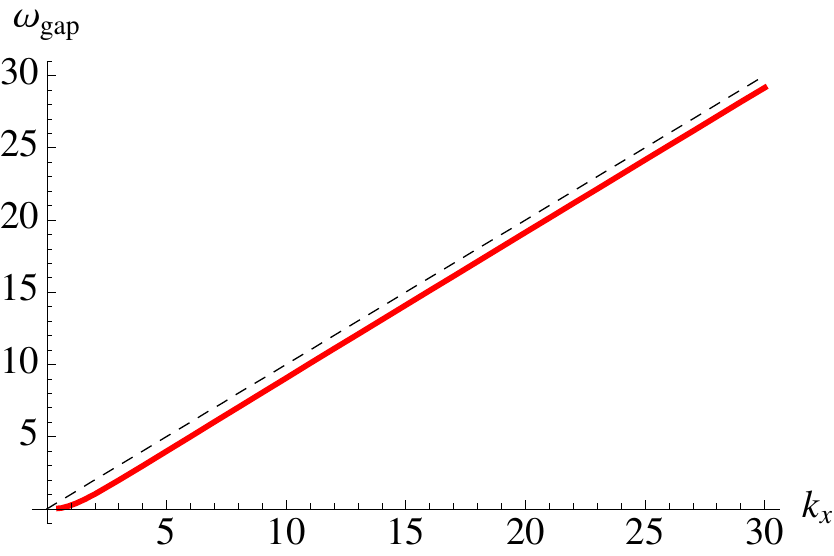}\label{fig:wgap0.35}}
\caption{\small Position of the edge of the gap for fixed $H$ as $k_x$ is increased.  The dashed line is the line of $\omega = k_x$.} \label{fig:gapbehavior}
\end{center}
\end{figure}
\item We find a peak for values of $\omega > \omega_{\mathrm{gap}}$.
  We refer to the position of the peak as $\omega_\ast$.  We show
  samples of the behavior of $\omega_\ast$ in figure
  \ref{fig:ridgebehavior}.  This is the ridge of
  ref.~\cite{Liu:2009dm}.  From figure \ref{fig:ridgebehavior}, one
  sees that as the magnetic field is increased, the ridge gets closer
  to the $w = k_x$ line, crosses it, and continues to approach it from
  above (see figure \ref{fig:ridge0.35}).  In addition, the ridge
  always remains above  the gap line (except when they meet).  We
  show  this in figure \ref{fig:ridge_gap_behavior}.  We also show
  an example of the large $k_x$ behavior for different magnetic fields
  in figure \ref{fig:large_kx_behavior}.
\begin{figure}[ht]
\begin{center}
\subfigure[$H=-0.1$]{\includegraphics[width=2.5in]{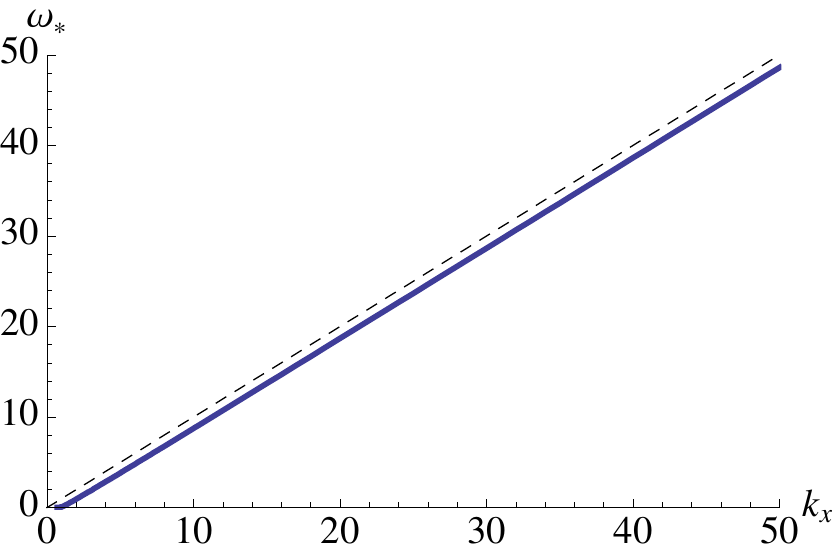}\label{fig:ridge0.1}} \hspace{0.5cm}
\subfigure[$H=-0.35$]{\includegraphics[width=2.5in]{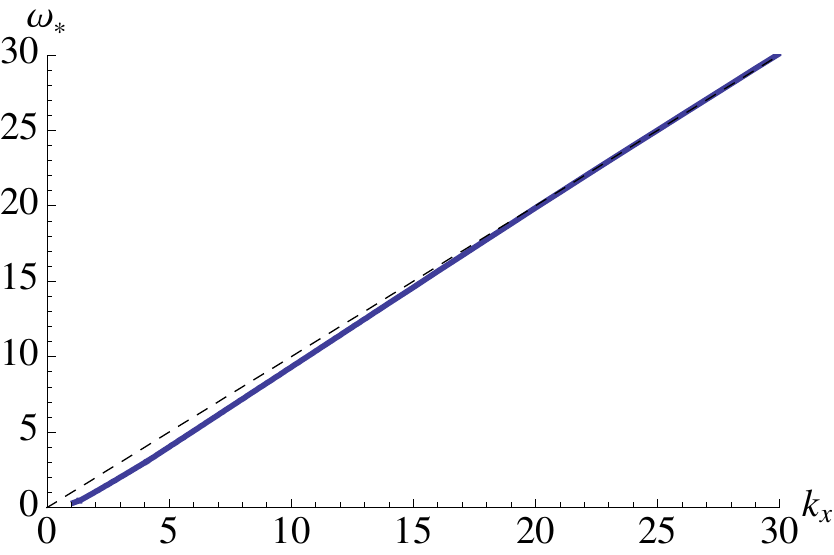}\label{fig:ridge0.35}}
\caption{\small Position of the peak for fixed $H$ as $k_x$ is increased.  The dashed line is the line of $\omega = k_x$.} \label{fig:ridgebehavior}
\end{center}
\end{figure}
\begin{figure}[ht]
\begin{center}
\subfigure[$H=-0.1$]{\includegraphics[width=2.5in]{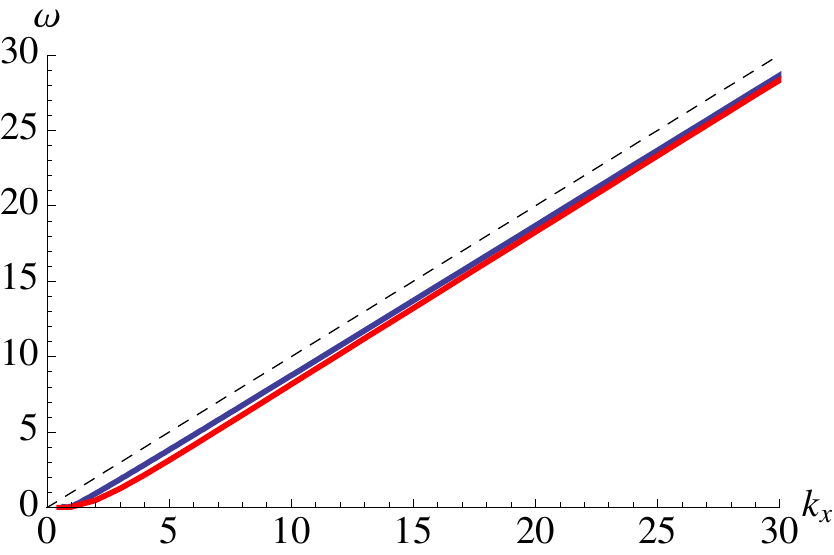}\label{fig:ridgegap0.1}} \hspace{0.5cm}
\subfigure[$H=-0.35$]{\includegraphics[width=2.5in]{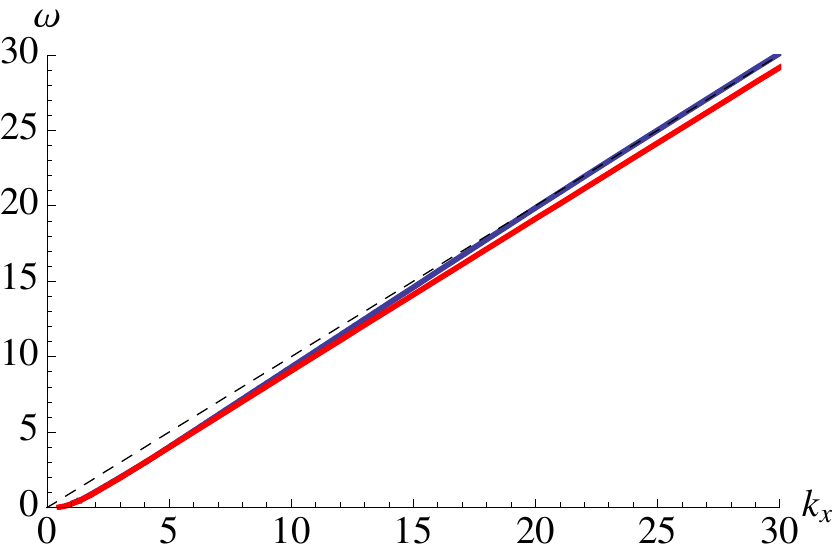}\label{fig:ridgegap0.35}}
\caption{\small Position of the peak and gap for fixed $H$ as $k_x$ is increased.  The dashed line is the line of $\omega = k_x$.} \label{fig:ridge_gap_behavior}
\end{center}
\end{figure}
 \begin{figure}[h] 
   \centering
   \includegraphics[width=2.5in]{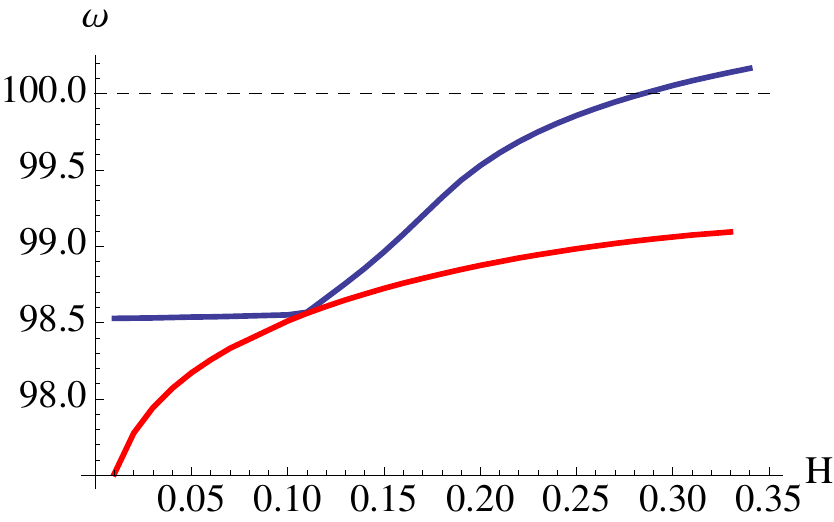} 
   \caption{\small Behavior of $\omega_{\mathrm{gap}}$ (red) and $\omega_\ast$ (blue) at fixed $k_x = 100$ }
   \label{fig:large_kx_behavior}
\end{figure}

\item For a fixed value of $H$, when $\omega_{\mathrm{gap}} =
  \omega_\ast$ at some $k_x=k_F$, we find a pole in Im$G_R(2,2)$.
  Notice that this means that a pole appears whenever the ridge curve
  touches the gap curve.  We find that the curves never cross but
  bounce off each other.  We show an example of such a pole in figure
  \ref{fig:pole_behavior}.  This is a quasiparticle peak. It indicates a
  Fermi surface at $k_F$.  Away from the poles, the
numerical results should be interpreted carefully since there one should consider a complex $\omega$ and not a purely
real~$\omega$.  This is relevant because in our numerics the poles
occur immediately after a region where the imaginary part of the
Green function is negative, which is not allowed if the theory is
unitary.  We believe that the proper treatment of the complex $\omega$
in these regions would resolve this.  
  
  The fact that $\omega>0$ tells us that the
  Fermi energy $E_F$ is greater than the $U(1)$ chemical potential
  $\mu$. This is quite natural: the physical magnetic field ${\cal
    H}=-2\alpha^2 H$ has lifted the energy of the charged particle, as
  earlier discussed. (Note that we see peaks for negative $\omega$ as
  well, but focus on positive $\omega$ for physical interpretation.)

  In particular, we note that the behavior is very reminiscent of the
  behavior of the full Green's function (that includes both the
  advanced and retarded Green's function) at the Fermi surface where
  both quasiparticles and quasiholes are observed. It is also
  important to note that there can be multiple (but we believe only a
  finite number) such peaks, at a given $H$, appearing at distinct
  values $(\omega_*, k_x)$.
 \begin{figure}[h] 
   \centering
   \includegraphics[width=2.5in]{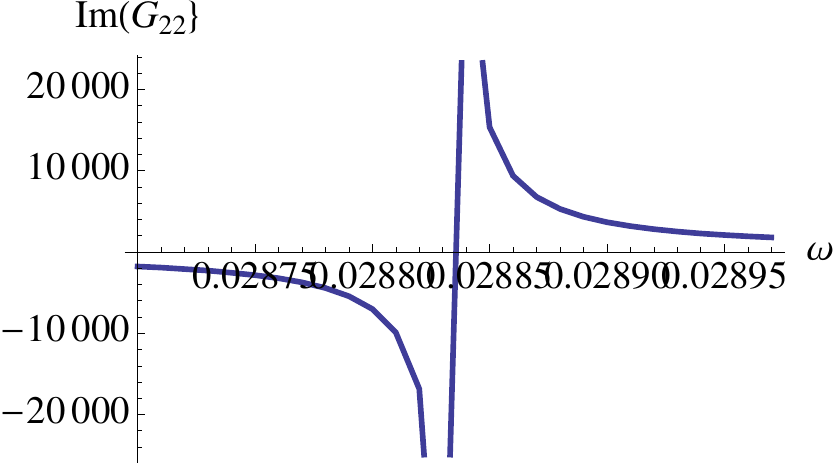} 
   \caption{\small Behavior of the pole in the Im$G_R(2,2)$ for $H=-0.1$ at $k_x$ =0.871847 }
   \label{fig:pole_behavior}
\end{figure}
\item There exists a maximum magnitude of the magnetic field at
  approximately $|H| \gtrsim 0.37$ above which we do not find any
  poles.  We notice that above this maximum, the ridge remains above
  the $w = k_x$ line for all values of $k_x$ and the gap remains below
  the line, preventing the two from meeting.  We explain in the
  following subsection what this maximum corresponds to.
\end{itemize}
\subsection{Dispersion of the Quasiparticle Peaks}
We proceed to study the scaling properties of the poles observed in
Im$(G_R(2,2)$.  We present samples of the results in figure
\ref{fig:w_scaling}.  We find that the behavior near the pole seems
very close to linear once $H\neq0$, if approaching from below or
above. Contrast this with the case of $H=0$ (discovered in
ref.\cite{Liu:2009dm}, recomputed here and displayed in
figure~\ref{fig:H=0_scaling}). However,   as one can see from the figure,
there is always a change in slope as we go from $k<k_F$ to $k>k_F$.
This suggests that there is perhaps curvature to the lines
infinitesimally close to $k_F$.  Therefore,  the system
is still in a non--Landau Fermi liquid regime  but with a scaling less than
that found for the $H=0$ case.  

Crucially, we see that as we increase the magnetic field, the slopes
of the two lines begin to approach each other.  Our results suggest
that when the slopes approach the limit of being equal, that is
exactly when we no longer find any poles.  This seems to be the
mechanism by which the system protects itself from becoming a simple (linear)
Landau Fermi liquid, as discussed in the introduction.
 \begin{figure}[ht]
\begin{center}
\subfigure[$H=-0.01$]{\includegraphics[width=2.5in]{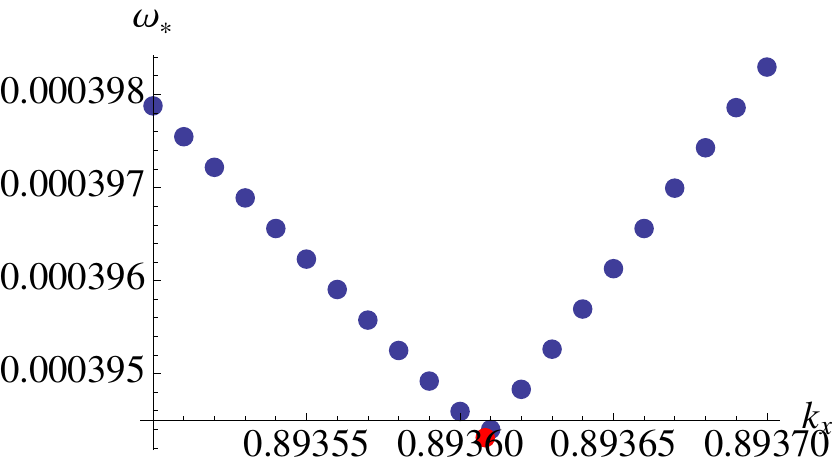}\label{fig:scaling_w_H=0.01}} \hspace{0.5cm}
\subfigure[$H=-0.1$]{\includegraphics[width=2.5in]{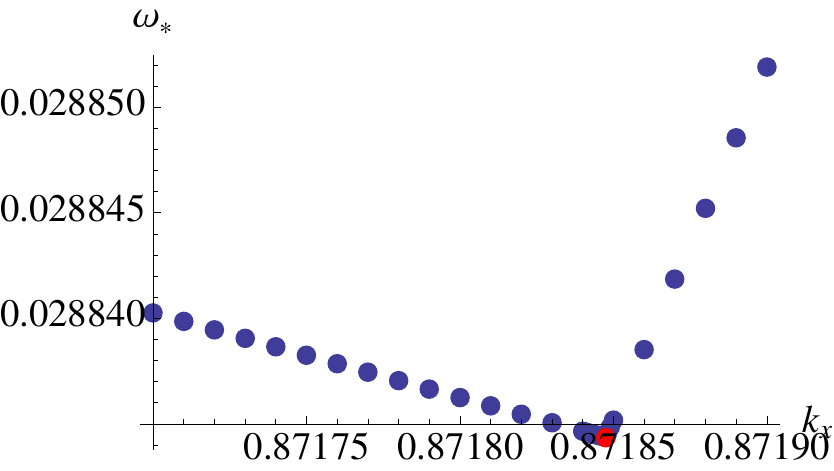}\label{fig:scaling_w_H=0.1}}  \hspace{0.5cm}
\subfigure[$H=-0.2$]{\includegraphics[width=2.5in]{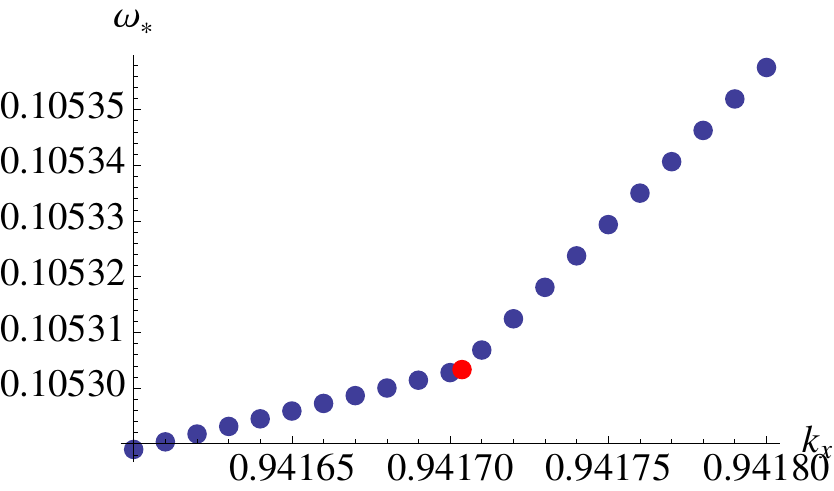}\label{fig:scaling_w_H=0.2}} \hspace{0.5cm}
\subfigure[$H=-0.35$]{\includegraphics[width=2.5in]{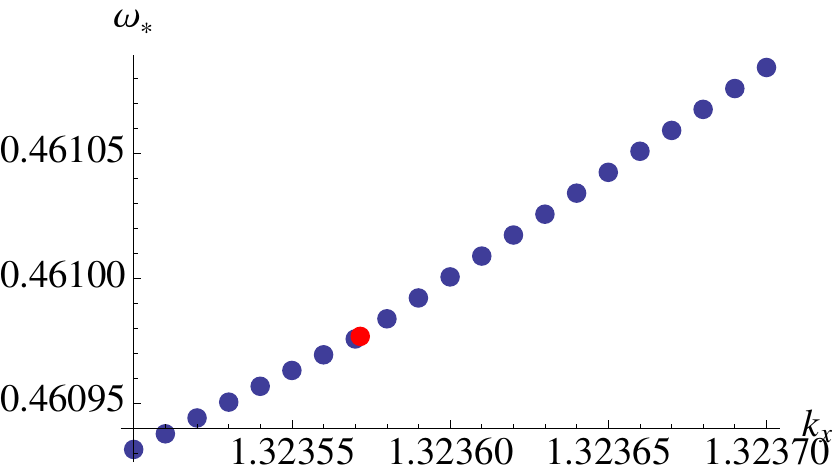}\label{fig:scaling_w_H=0.35}} 
   \caption{\small Scaling behavior of $\omega_*$.  The red dot is where the pole is located, which corresponds to where the two lines intersect.}  \label{fig:w_scaling}
   \end{center}
\end{figure}

In addition, we note that when there are are multiple poles present
(for a given $H$), the dispersion behavior of the poles may not be the
same.  We present an example of this in figure
\ref{fig:w_scaling_fixed_H}.  The results suggest that the further
along the ridge one finds a pole, the more the behavior of the pole
approaches that of a Landau Fermi liquid ({\it i.e.,} linear dispersion).  This
suggests why there may only be a finite number of poles for a fixed
magnetic field, since as we saw earlier, beyond a certain $H$, or far
enough along the ridge, the ridge and gap no longer cross.
 \begin{figure}[ht]
\begin{center}
\subfigure[First Pole]{\includegraphics[width=2.5in]{w_pole_H=0_35_RHS}\label{fig:scaling_w_H=0.35_v1}} \hspace{0.5cm}
\subfigure[Second Pole]{\includegraphics[width=2.5in]{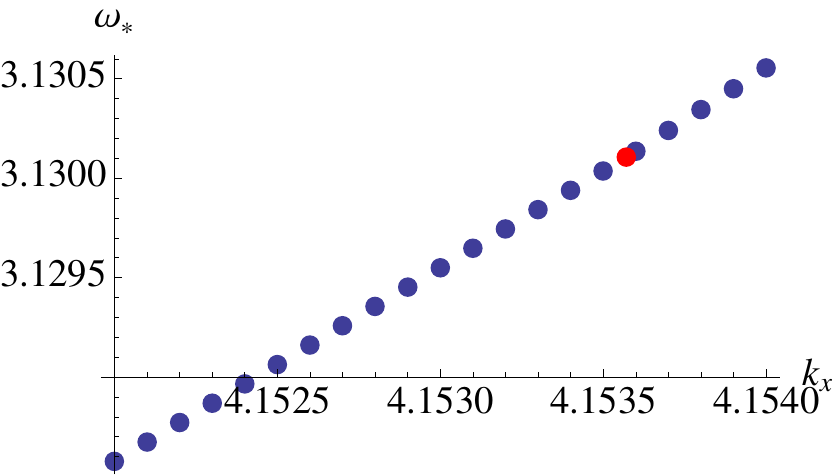}\label{fig:scaling_w_H=0.35_v2}} 
   \caption{\small Scaling behavior of $\omega_*$ for two poles at fixed magnetic field $H=-0.35$.  The red dot is where the pole is located, which corresponds to the two lines intersect.}  \label{fig:w_scaling_fixed_H}
   \end{center}
\end{figure}

\section{Conclusions}

We've found a rich set of physical results (the deformation of
dispersion characteristics, and the discrete Fermi levels) from our
holographic studies of the Fermi surface and quasiparticle spectrum in
a background magnetic field. We expect that there is even more rich
physics to be found from these systems. It is very exciting that some
of these phenomena seem akin to the sorts of strongly coupled physics
that are experimentally accessible, including with background magnetic
fields as a probe of physics.

Our exploration work was primarily numerical. A precise analytic
relation between the value of the magnetic field and the nature of the
dispersion of the peaks would be a valuable result. However, there is
an apparent obstruction to doing the obvious generalization of the
analysis of ref.\cite{Faulkner:2009wj} that exploits the presence of
the AdS$_2\times\mathbb{R}^2$ region. There, an expansion about the
${\omega}=0$ point, where the quasiparticle peak appears, representing
a Fermi surface with Fermi energy $E_F$ equal to the $U(1)$ chemical
potential $\mu$.

Things are different in our case. While it is trivial to show that
there is again an AdS$_2\times\mathbb{R}^2$ region near the horizon of
the dyon (rather nicely, the electric field is entirely in the AdS$_2$
and the magnetic field is entirely threading the $\mathbb{R}^2$), this
is not enough. In the presence of magnetic field, the energy of the
system gets lifted. The lowest available level at which we find a
quasiparticle peak, for a given ${\cal H}$ has Fermi energy $E_F$
greater than the $U(1)$ chemical potential $\mu$, and so occurs away
from the ${\omega=0}$ point that was the focus of
refs.\cite{Liu:2009dm,Faulkner:2009wj}. There is no good analytical
guide (so far) as to where in $\omega$ the peaks will arise, and so we
search for them numerically. An analytical characterization of exactly
how the presence of ${\cal H}$ affects the dispersion relation will
have to await future work.

\section*{Acknowledgments}We would like to thank Arnab Kundu and
Mohammad Edalati for conversations that led to this project. CVJ thanks
Ruth Andrew Ellenson for an insightful remark  and the Aspen Center for
Physics for a stimulating working atmosphere while some of this work was carried out.  This work was supported by the US Department of Energy.


\providecommand{\href}[2]{#2}\begingroup\raggedright\endgroup

\end{document}